\documentstyle[aps,epsfig,twocolumn]{revtex}

\newcommand{\eeq}{\end{equation}}
\newcommand{\beq}{\begin{equation}}
\newcommand{\nuq}[1]{\label{#1} \eeq}
\newcommand{\veps}{\varepsilon}

\begin{document}

\title{ 
Quantum Algorithmic Integrability: \\
The Metaphor of Polygonal Billiards.
}
\author{ Giorgio Mantica \\
         International Center for the Study of Dynamical Systems \\
 Universit\`a della Insubria, via Lucini 3, 22100 COMO ITALY \\
 INFM-INFN unit\`a di Milano - giorgio@fis.unico.it }
\maketitle

\begin{abstract}
An elementary application of Algorithmic Complexity Theory to the 
polygonal approximations of curved billiards--integrable
and chaotic--unveils the equivalence 
of this problem to the procedure of
quantization of classical systems: 
the scaling relations for the 
average complexity of symbolic trajectories 
are formally the same as those governing
the semi-classical limit of quantum systems.
Two cases--the circle, and the stadium--are examined in
detail, and are presented as paradigms. \\
{\em PACS 05.45.-a,  05.45.Mt,  89.70.+c} \\
{\em 1991 Math. Subj. Class. 70K50, 94A15, 81S99} 
\end{abstract}

\begin{center}
{\em Dedicated to Boris V. Chirikov for his seventieth birthday}
\end{center}
\section{Introduction}

{\em Chaos} is certainly the most significant concept that
has issued from the theory of dynamical systems and yet 
its true meaning, most concisely and universally encompassed in the 
equation {\em chaos = deterministic randomness} has
not been fully adopted in the literature and in
the scientific community.
This is somehow paradoxical, for even in popular magazines the idea
has spread that chaos theory might be considered
the third scientific revolution of the century--after 
relativity and quantum mechanics.
In my opinion, two are the
main reasons behind this failure:
firstly, the information-theoretical concepts implied
in the notion of  {\em deterministic randomness} are unfamiliar to most
scientists; secondly, for the vast majority of physicists
the ``true'' mechanics is not classical--where chaos is
commonly found--but quantum, where chaos is, quite significantly
as we shall see, absent.

A clash is implied in this last statement: if chaos is absent in quantum
mechanics, shouldn't it be also in classical mechanics, which is 
just the limiting case of the former ?  
This clash has lead people to draw all possible sorts
of conclusions. Many have claimed that classical chaos
must imply quantum chaos, via the {\em correspondence principle}
\cite{cp},
while others on the same basis
have pretended that classical chaos theory should be derived from
quantum dynamics. 
Others yet
have argued that being chaos absent in the
quantum mechanical theory of nature, it should be absent 
in nature altogether, and in particular at the macroscopic level,
and so farewell classical chaos. 
Taken at face value, this last statement implies
a logical inconsistency, for quantum mechanics is just a 
theory, a {\em description}
of nature. Yet, many physicists consider it a {\em very good}
description in all respects, including chaos or its lack. 
The fact remains that chaos in nature 
is an undisputed reality, quite well described 
by classical dynamics.
In previous works we have put forward the idea that 
the gap between the two mechanics, classical and quantum,
is wider than can be naively expected from the correspondence
principle 
\cite{physd,copen,ilg,amj}.

In this paper I shall show that a similar situation is
met in the dynamics of classical billiards: here, {\em 
rational polygonal} billiards play the r\^ole of quantum systems,
whose ``classical limit'' are {\em curved} billiards,
to which they tend geometrically as the number of polygonal sides 
increases indefinitely.
Indeed, while it has been since long recognized that the former are
models of non-chaotic behaviour \cite{ford1,zeml,sina,gutk,riche}, 
one can use them to approximate chaotic curved billiard tables to
arbitrary precision, and ask what happens then of the character
of their motion. Even more:
recently some observations
of positive ``effective'' Lyapunov numbers 
in polygonal billiards
have been published \cite{ford2}, 
and a paradox of the same flavor as
``quantum chaos'' seems to arise.
To resolve this paradox,
in this paper I shall introduce an elementary, ``physical''
version of algorithmic complexity theory. In developing this 
theory, it will become clear
that the procedure of approximating 
curved billiards by polygons is 
quite analogous to that of
quantizing classical systems:
understanding the complexity of the motion in polygons
can then  be profitably used to
clarify the issues
involved in the other, more important problem.

Our arguments are organized as follows: in the next section we
review the fundamental pieces of  information on integrable and chaotic
billiards, and the notion of {\em algorithmic integrability}.
To adapt algorithmic complexity to
physical purposes, 
in Sect. III a simple {\em coding} of trajectories
in billiards is introduced, which translates these
into symbolic sequences. In Sect. IV the attention is focused on the
{\em scaling} of the complexity of these sequences with respect to
time, within certain time intervals:
this leads to the concept of {\em randomness (or order) within a range}.
The theory is immediately applied to the case at hand: Sect. V 
introduces
two families of billiard tables: the circle
(an integrable system), the stadium (a fully chaotic one), and their 
rational
polygonal approximations, which, by analogy to quantum mechanics
we also call {\em polygonal quantizations}.
Rather than studying orbital complexity directly, 
I introduce here the concept of {\em average coding length}, which 
has a clear physical meaning, and can be used to estimate the former.
The case of circle quantizations is then studied in Sect. VI, where 
the paradox presented above is resolved. The stadium
billiard is put to the same test
in Sect. VII, and the problem of {\em correspondence} is addressed.
Finally, I shall try to draw some general conclusions
asking a challenging theoretical question.

It is a pleasure and a honour for me to 
to dedicate these reflections to Boris Chirikov
for his seventieth birthday as my 
teacher and colleague Joe Ford would have certainly
done, had he been here now.

\section{Chaos and Complexity in Billiards} 

Billiards are perhaps the dynamical
systems that require less of an introduction,
and just a few formal definitions are necessary.
A {\em billiard  table} $B$ is a bounded, connected domain of the plane,
with piece-wise smooth boundary. 
An {\em ideal billiard} in $B$ is the dynamical system originating from the
uniform motion of a point particle--a ball--inside
$B$, with elastic reflections at the boundary, following
the familiar law of {\em angle of incidence equal angle
of reflection}.

In many games that can be played on a billiard, 
attention is paid at such reflections:
one discretizes time (the integer $n$ meaning the time of
the $n$-th rebound) and considers
the subset ${\cal S}$ of the tangent space of $B$,
consisting of unit tangent vectors attached at boundary points,
pointing inside $B$.
${\cal S}$ can be easily parameterized by the pair $(l,\phi)$,
where $l$ is the arc length along the boundary, 
and  $\phi$ is the angle between the unit vector and 
the inner normal to the boundary, 
$-\frac{\pi}{2} < \phi < \frac{\pi}{2}  $. 
In so doing, dynamics is a function $T$ from ${\cal S}$ to itself,
that maps the bounce occurring at $l_{n-1}$, with ``exit angle''
$\phi_{n-1}$, into the new collision point $l_n$, and exit
angle $\phi_n$:
\beq
    (l_n,\phi_n) = T (l_{n-1},\phi_{n-1}).
\nuq{bil1} 

Since this mapping preserves the canonical measure
$
   d \mu = \cos \phi \; d l \; d \phi ,
$
billiards are among the simplest and
most successful examples of conservative dynamics. 
But perhaps
they mostly owe their success to the fact 
that a member of their family can be found at virtually
all levels in the famous ergodic hierarchy: there are integrable
ones--the {\em circle}--as well as $K$--the {\em stadium}, which is therefore
also ergodic, and mixing. 
Donald Ornstein has shown (with Gallavotti \cite{galla})
that billiards can also be Bernoulli,
and has conjectured (with Weiss \cite{orns}) that chaos in nature is mostly 
of this type: not without reason,
we can say that billiards have served to shape 
our view of reality. In a sense, the present paper aims at
the same ambitious goal.

An interesting sub-class will be studied here,
which covers part of the ergodic hierarchy, but falls short of
producing chaotic representatives: the polygonal billiards
\cite{gutk}.
Sinai has indeed proven \cite{sina} that these billiards have 
null metric entropy, and Ford has termed polygonal billiards with 
rational angles
{\em algorithmically} integrable ({\em A}) \cite{ford0},
where the letter {\em A}
was also intended to honor the memory
of V.M. Alekseev, and his work 
\cite{aleks} which showed that orbits of null entropy systems have
null algorithmic complexity.
The terminology  makes it evident that a shift in perspective
has taken place:
while the ergodic hierarchy
is concerned with statistical properties of ensembles
of orbits, algorithmic  theory deals with a
new object, the complexity of the 
description of the motion, which will be the basis of
our investigation \cite{remo}.

Seminal work \cite{geppo} on {\em A}-integrable billiards is 
Zemlyakov and Katok's \cite{zeml} study of polygons whose
vertex angles are all rational multiples of $\pi$:
it shows that these billiards
satisfy the conditions for integrability except
for the effect of vertices. 
Rationality of the angles provides a second constant of
the motion (the angle of reflection $\phi$ times a
suitable integer multiple of $\pi$), but {\em splitting} of trajectories
heading on a vertex provides the {\em error}
\cite{hobs}
which prevents the system from being integrable.
Notwithstanding these errors trajectories 
are still computable, in the sense that effective algorithms 
can be devised, in such a 
way that the number of informational
bits in the output (the trajectory) is much greater than the number 
of informational bits in the input (the algorithm 
{\em plus} the initial condition of the motion).
Echardt {\em et al} in \cite{ford1}
call such trajectories {\em algorithmically meaningful}.
Therefore, {\em A}-integrable systems are computationally
akin to Liouville-Arnol'd integrable ones
\cite{riche2}. We shall 
come back to these concepts in the following.

A few years after ref. \cite{ford1} Ford and other co-workers returned 
to the theme of rational billiards, presenting a seemingly  
different set of conclusions \cite{ford2}: they examined the rate
of divergence of nearby trajectories and found that
this rate is exponential,
even for rational billiards. The key factor in their derivation
is the fact that nearby trajectories differ at time zero
by a finite, fixed amount, and they are re-initialised to this
fixed amount at each iteration of the Benettin-Strelcyn algorithm
which computes Lyapunov numbers. In the mind of these
authors, this upper bound to precision stands
for the human limitation against the dogma of infinite
precision. If we take this limitation
into account, they claim, trajectories effectively
live on a multi-sheeted surface to which splitting
at  rational vertices 
gives an average negative curvature: the resulting
motion is, practically speaking, chaotic.
The first aim of this paper is to put order in these conflicting
observations, by utilizing a {\em scaling} approach to
algorithmic complexity theory. But first, let us 
define the rules of our game.

\section{Symbolic Sequences in Billiards: a Game }

\begin{figure}[htbp]
\centering\epsfig{file=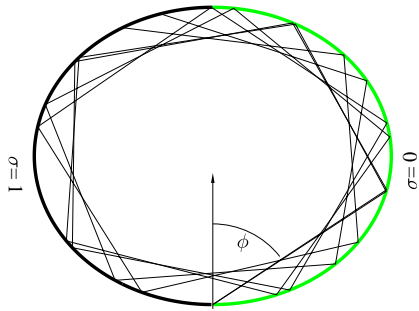,width=1.0\linewidth}
\caption{The initial
portions of two trajectories in a circular billiard table,
characterized by slightly different initial angles $\phi$.
Their common initial position is at the bottom of the table.
Also shown is a coding of rebounds with two symbols,
$\sigma = 0$ (light portion of the boundary) and 
$\sigma = 1$ (dark portion).
}
\protect\label{cerchio}
\end{figure}

\begin{figure}[htbp]
\centering\epsfig{file=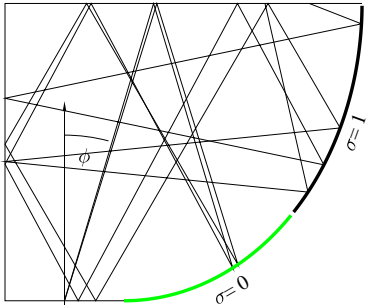,width=1.0\linewidth}
\caption{Same as Fig. 1, now for the quarter stadium billiard table.
Notice that rebounds are coded only on the circular part of the boundary,
by $\sigma=0$ (light thick portion of the boundary) and by 
$\sigma=1$ (dark thick portion).
}
\protect\label{stadio}
\end{figure}

The billiard in a circle
(Fig. \ref{cerchio})
is a noticeable example of Liouville-Arnol'd
({\em L-A}) integrability: the second, smooth integral of the motion 
being the angular momentum with respect to the center.
Cutting the circle in two equal pieces, and inserting 
a rectangular strip in between the two halves gives
rise to a fully chaotic billiard: the stadium \cite{bunny}. 
We shall get rid of all symmetries in this
geometrical figure, and study the {\em quarter stadium}
(Fig. \ref{stadio}).
Let us now replace the circular sides in both billiards by
a polygonal approximation
with equal sides: it is apparent that this can be
done so to form a rational billiard.
In both these cases, Vega, Uzer and Ford have found exponentially
divergent trajectories, within their approximation scheme of course.
In the following, I shall show
that the two cases are different indeed.
To do this, I first need to introduce a {\em symbolic coding}
of this problem.

How to {\em code} a dynamical object into a symbolic sequence
is something which follows either from physical insight, 
or practical convenience, or mathematical efficacy.
In our case, we shall elect to code trajectories according
to the bounce coordinate $l$ alone: for this, we shall
assume that the circular parts of the boundary (or their
polygonal approximations) are divided into a finite number,
$S$, 
of equal regions, each of which corresponds to a symbol,
$\sigma$, which can be taken to be a natural number from
$0$ to $S-1$. We also decide that bounces
on the straight segments of the stadium will not be registered.
This coding is indicated in Figures
\ref{cerchio} and \ref{stadio}, for $S=2$.
In a billiard game like this,
we may think of putting detectors all around the 
boundaries, which are set off whenever the ball hits them. 
In the polygons,
nearby sides can be connected to the same detector,
so that the total number of output channels is $S$, also
when $M$--the number of sides--is much larger than $S$,
and is allowed to increase, while keeping $S$ fixed,
as we shall do in the following.
The history of a trajectory is
then coded in the record $\sigma_1,\sigma_2,\ldots$ of
boundary reflections.

Now, let us play a funny game:
the ball is initially set still at a fixed
point, and a test player--chosen 
appropriately among our pals--can
aim it by hitting it properly by the conventional cue.
As the departing angle $\phi$ is varied
different trajectories are initiated, and
different symbolic sequences $\{\sigma_j\}$ are recorded.
Let us now ask the player to {\em do it twice},
that is, to aim the ball a second time so that 
the sequence of bounces he had obtained in the first
shot is repeated \cite{notgam}. It comes as no surprise
that he will not be able to set the initial
angle $\phi$ to {\em exactly} the same value in both
tries: the difference in this quantities causes
the two resulting symbolic trajectories to agree only over a finite
time-span. 
Trajectories of this kind are pictured in Figures
\ref{cerchio} and \ref{stadio}. 
But how is this sport related 
to algorithmic complexity and to
our problem ?

\section{Deterministic Randomness of Finite 
Trajectories}

In the light of the previous section,
a billiard can be thought of as a computer program, that is, 
a Turing machine capable of taking a binary sequence 
$p_1,\ldots,p_L$--the program, to which we shall
momentarily return--as input,
and producing a symbolic sequence, $\sigma_1,\ldots,\sigma_N$,
the output. Recall now that
the algorithmic complexity $K(N)$ of a sequence $\{\sigma_j\}$, 
for $j = 1, \ldots, N$ is defined very roughly as the length of the 
{\em shortest} computer program capable of outputting
the sequence, and stopping afterwards \cite{kolm,chai}.  
We can say that
the complexity of a sequence is the length of its shortest definition:
\beq
    p_1, \ldots, p_L
\stackrel{\mbox{outputs}}{\Longrightarrow }
    \sigma_1,\ldots,\sigma_N \;
\mbox{ implies} \; K(N) \leq L.
\nuq{comple1}                  
Algorithmic complexity theory teaches us that most sequences--in probabilistic
sense--of length $N$ have complexity close to maximal, i.e.
$N$. Moreover, in the limit of increasing $N$, almost all
of them are {\em random}, in the sense that
their complexity $K(N)$ grows as $N$
\cite{lof}.
At the same time, {\em computable} sequences exist,
for which $K(N)$ is much less than $N$, and
grows less than linearly.
We apply now this theory to the dynamical sequences 
constructed in the previous section.
Let us call their complexity
$K(N,\phi)$, where we have explicitly indicated
that this number may depend on the initial condition of
the motion.
We shall then follow Alekseev \cite{aleks}, Chirikov \cite{treotto}, and Ford
\cite{fordm1},
and identify 
{\em order} with {\em computability}, and 
{\em chaos} with {\em randomness}.

The central issue is then to find a computer code to output the
desired sequence. For our dynamical sequences,
$\sigma_1, \ldots, \sigma_N$, 
this can be obtained in any abstract language by an encoding of:
a) the geometrical rules of the game, 
which require
a fixed number of bits,
$C_{\mbox{\scriptsize machine}}$ (which depends only on the machine
on which the rules are coded),
{\em plus} b) the instructions set for fixing
the billiard boundaries, of length
$C_{\mbox{\scriptsize boundary}}$,
{\em plus} c) the specification of $N$
{\em and} d) of a certain number of digits of $\phi$. 
Accordingly,
the length $L(N,\phi)$ of this program can be estimated as 
\beq
     L(N,\phi) \simeq 
     C_{\mbox{\scriptsize machine}}  +
     C_{\mbox{\scriptsize boundary}} + 
    \log_2 N + \Lambda(N,\phi),
\nuq{big1}
where the function $\Lambda (N,\phi)$
is defined as the {\em number of bits of $\phi$ necessary
and sufficient to determine the first $N$ symbols 
in the sequence $\sigma$}. 
This function can serve
to estimate the complexity $K(N,\phi)$ in its
most important aspect: the $N$ dependence. 
For it is clear that
the opposed behaviors presented above--random
and computable--are
of deep physical significance: if in a range
$N_l \leq N \leq N_u$ 
the function $\Lambda (N,\phi)$ grows linearly,
$
    \Lambda (N,\phi) \sim \lambda N ,
$
the complexity $K(N,\phi)$ has the same leading behaviour,
and the sequence $\{\sigma_j\}$ should  duly be termed random,
or chaotic.
When, in a similar interval, $\Lambda (N,\phi)$ grows less 
than linearly, $\{\sigma_j\}$ should be called computable, or ordered.
We therefore introduce the notion of order (and randomness) 
{\em within a range}. The physical significance of such order (or randomness)
will then be proportional to the importance of such range.

We have emphasized this {\em scaling} approach in
\cite{copen} to rebut a common objection to the
application of complexity theory to finite dynamical sequences
\cite{notfin}.
The rationale behind this idea is evident, and is 
quite similar to that adopted in the independent theory
of computational time complexity: not so much the 
value of complexity is relevant, but the way it increases
as the problem size grows, for this may render
it quickly unfeasible.
In fact,
there is no point in practicing for our billiard
player if he is playing a chaotic billiard:
for each additional bounce
he wants to set correctly his aiming precision in the initial
angle $\phi$ must increase geometrically.
Out of metaphor, when the billiard is a system 
whose symbolic dynamics we want to predict, to obtain 
a linear increase in forecast precision we must 
exponentially increase the accuracy in the 
initial conditions \cite{brudn}. 
In most instances this demands an
exponential increase of economical resources.

\section{The Circle and the Stadium--Polygonal Quantization} 

To draw a parallel with quantum mechanics \cite{balla}
suppose now that Nature, by means of our observations,
clearly shows that {\em curved} billiard boundaries are
a mathematical idealization and that {\em physically}
we can just have {\em polygonal} billiards 
with an arbitrary but  finite number of sides, $M$.
These sides will be inscribed in the circle and in the
stadium, and 
I shall call the resulting polygonal billiards 
a {\em quantization} of the curved ones. In this context,
$M$ is a crucial quantity,
which plays the r\^ole of a semi-classical parameter.
Clearly, when we let $M$ go to infinity
we regain geometrically the original table.
The question is, will we obtain the same dynamics ?

To answer this question in a complete, algorithmic sense,
let us consider again the coding introduced in 
Sect. III, and let us study the function 
$\Lambda(N,\phi)$ defined in Sect. IV, or rather its inverse,
$N(\Lambda,\phi)$, which quantifies the number $N$
of symbols $\sigma_j$ which can be predicted knowing
$\Lambda$ digits of $\phi$.
It is clear that order (or randomness)
of a dynamical sequence can also be inferred from the
study of $N(\Lambda,\phi)$. Moreover,
this function has a transparent physical meaning:
if we let $\veps :=  2^{-\Lambda}$,
in this new independent variable the function
$\tilde N(\veps,\phi) := N(-\log_2 \veps,\phi)$ 
can be called the length of the codable 
trajectory under {\em uncertainty} $\veps$.
Let us go back to the player example.
He is only capable of setting the angle $\phi$ with a
certain error $\veps$. 
Under these circumstances, only  finitely many dynamical symbols
will typically coincide in the initial portions of
a series of two shots $\sigma$ and $\sigma'$: their number 
is $N(\Lambda,\phi)$,
\beq
  N(\Lambda,\phi) = k \; \mbox{ {\bf if}  } \;
   \sigma_1 = \sigma'_1, \ldots ,
   \sigma_k = \sigma'_k,
   \sigma_{k+1} \neq   \sigma'_{k+1}.
\nuq{diffe}                                   
Following \cite{discr} we also say that
$\tilde N(\veps,\phi)$ is the {\em first error time},
when any two initially $\veps$-close trajectories 
start to have different symbolic sequences.

Numerically this function is rather straightforward to
compute, and its theoretical analysis can
be carried out in full detail (see the Appendix and below).  It is plotted
in Fig. \ref{cer2a.gle} 
versus $\phi$ at fixed uncertainty $\veps$, for the $M=8$ quantization
of the circle (a regular octagonal billiard).
One observes that low values,
like that seen at $\phi_1$,
are associated
with trajectories heading very early 
in history towards a vertex, where 
a {\em symbolic error} may occur \cite{discr,ford1}.
Large values of $N(\Lambda,\phi)$ are found when
this happens much later: $\phi_2$ is  
associated with a periodic trajectory which stays
far away from the vertices \cite{perio}.
 
It can be shown that the behaviour of $N$ at the periodicity values
is at the root of the dynamical
properties of the system. In fact,
let the ideal, frictionless ball run 
for an infinite time.
A coding function $F$ of the initial angle $\phi$
can be defined as
\beq
    F(\phi) = \sum_{j=1}^\infty \sigma_j S^{-j}.
\nuq{code1}
This function represents the translation dictionary between
the trivial code--the infinite sequence of digits of $\phi$,
and the dynamical code--the infinite sequences of bounces.
In chaotic systems, like the motion over surfaces of constant
negative curvature, and the anisotropic Kepler problem,
the relation between the properties of $N$ and 
the function $F$ can be fully exposed. This relation 
justifies the multi-fractal properties 
of the coding function $F$, 
which were originally investigated 
by Gutzwiller and Mandelbrot \cite{guzzi,gioguz}.

\begin{figure}[htbp]
\centering\epsfig{file=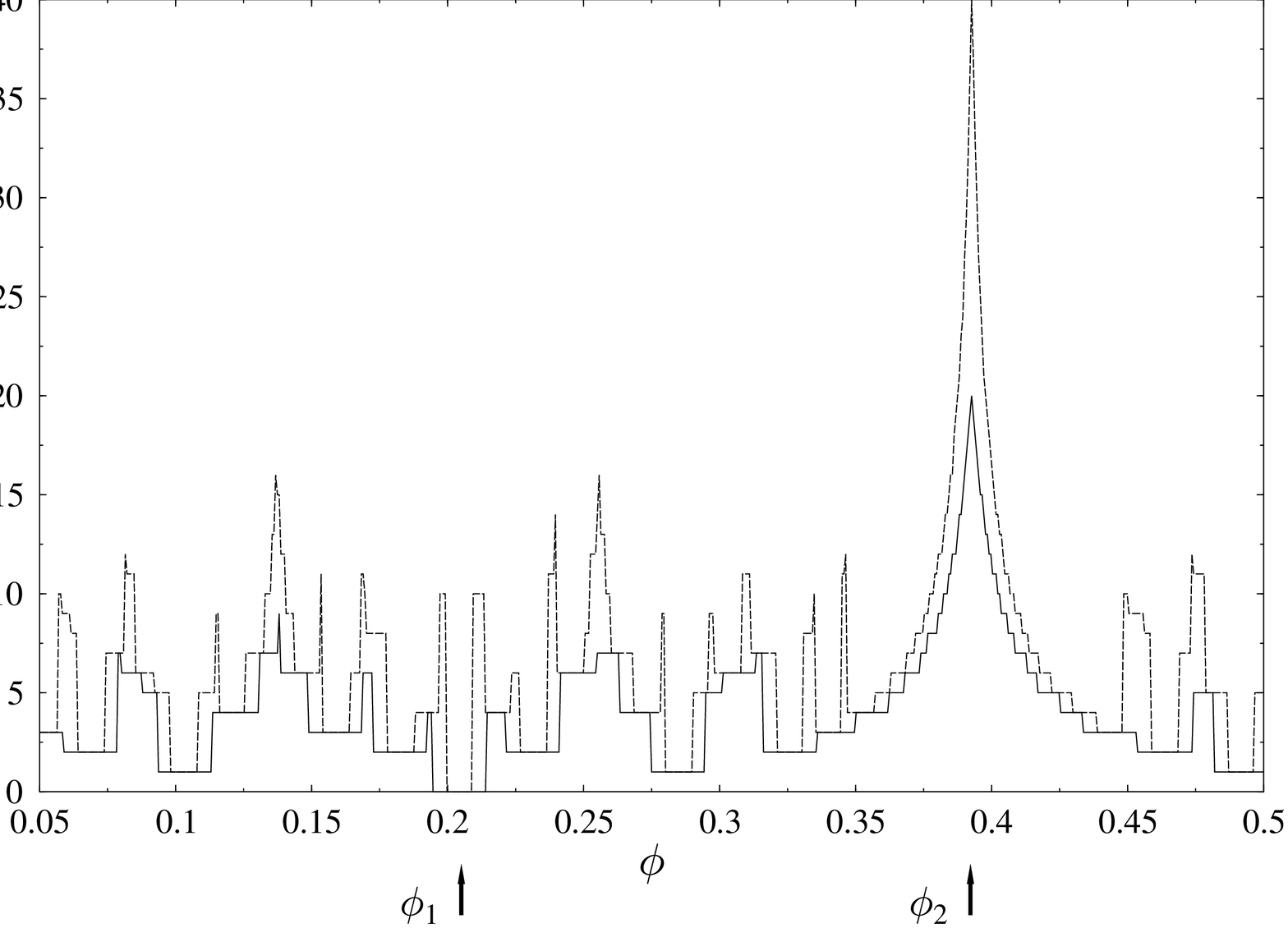,width=0.9\linewidth}  
\caption{Coding length $N(\Lambda,\phi)$ vs $\phi$ for an
octagonal billiard, with $S=8$. The starting point is at the center 
of one of the sides. Two values of $\veps$ are used:
$\veps = .01$ (bottom continuous line) and $\veps=.005$ (top 
broken line).
The arrows mark the values $\phi_1$ and $\phi_2$ described in the text.
}  
\protect\label{cer2a.gle}
\end{figure}

In player's terms, the angle $\phi_2$ in Fig. \ref{cer2a.gle} is an
{\em easy shot}, and $\phi_1$ a {\em tough} one:
we find then convenient
to average over initial angles, so 
defining the average coding length $A(\veps)$:
\beq
    A(\veps) := \int N(-\log_2 \veps,\phi) \; \cos \phi \; d \phi ,
\nuq{media}
which will become our main indicator 
of the complexity of the motion. 
A canonical average over the full phase space can 
be also defined and computed, with quantitatively similar results
to those we are going to expose. 
Let us pause no more,
and compare the behavior of this quantity in the circle,
the stadium, and their polygonal quantizations.

\section{Orbital complexity in integrable and 
polygonal billiards}

\begin{figure}[htbp]
\centering\epsfig{file=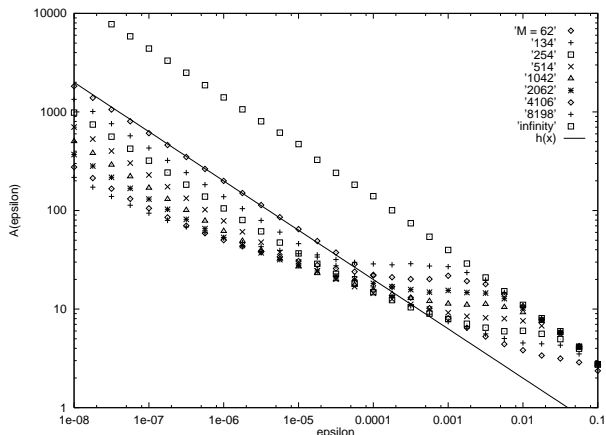,width=0.7\linewidth,angle=270}  
\caption{Average coding length $A_M(\veps)$ vs $\veps$ for 
the circle and its polygonal quantizations, with $S=2$. 
The values of $M$ are chosen to be twice a prime, and
roughly geometrically increasing.
The continuous line is the second formula in eq. (\ref{forcer})
with $M=62$.  }  
\protect\label{cer2.gle}
\end{figure}

We have now set the stage for the examination of the average 
orbital complexity in our billiards. 
We start from the case of the circle and its polygonal
approximations:
before resorting to
exact analysis, let us have a look at the numerical data.
Fig. \ref{cer2.gle} 
plots $A(\veps)$ versus $\veps$.
The scale
is doubly logarithmic: 
according to eq. (\ref{big1})
the power-law 
behavior of these curves for
small $\veps$ is a manifestation of the ordered character of the
motion, which is 
well assessed in the literature
\cite{sina}, \cite{aleks}, \cite{ford1}.

Yet, the physical picture evident in Fig.  \ref{cer2.gle}
is much richer.
Firstly, as $M$ grows at fixed (large)
$\veps$, $A_M(\veps)$ tends to $A_\infty(\veps)$, where the 
label $M$ is self-explanatory, the earlier the larger the
value of $\veps$. This is indeed clear: at large $\veps$ the coding
length is short, and differences between  polygonal boundaries
with large $M$ and the circle are not significant. 
But if we turn now
our attention to the left part of Fig.  \ref{cer2.gle}, where relatively 
smaller values of $\veps$ are plotted, a less expected phenomenon
appears: increasing $M$ at fixed $\veps$  the difference between  
$A_M(\veps)$ and $A_\infty(\veps)$ seems to {\em grow} rather than
vanish. Moreover, this discrepancy is brought about by a 
{\em decreasing} $A_M(\veps)$. 
Recall that 
$A_M(\veps)$ is the average of the inverse function of 
$\Lambda(N,\phi)$ (if you rotate the graph clockwise by ninety degrees,
$N$ appears plotted on
the horizontal axis and $\veps = 2^{-\Lambda(N)}$ on the vertical):
in this region the complexity of a trajectory
of fixed length $N=A_M$ {\em grows} 
when $M$ is increased. This is the phenomenon
observed by Vega, Uzer and Ford: polygonal
billiards have their own brand of instability, generated by splitting
of trajectories at vertices. The more the vertices, the larger
the Lyapunov numbers computed in \cite{ford2}. 
The paradox presented in Sect. II is so exposed.

But we are now equipped to resolve it: observe first 
that the increase in complexity obtained by 
raising $M$ at fixed $\veps$ is only temporary: if we 
keep going, we find that 
$A_M(\veps)$ finds
a minimum, and then inverts its course
to reach quickly
(we shall come back on the rate of this convergence later on)
the circle value $A_\infty$ \cite{notve}. 
Furthermore, a deeper argument must be made: 
sitting at fixed $\veps$ is not 
a proper thing to do, not even in the presence of human limitations to
finite precision. In fact,
we cannot increase
precision {\em indefinitely}, but we certainly can over 
a finite, physically reasonable range. For instance, this
may be dictated by the computational power of the 
machine on which Fig. \ref{cer2.gle} has been computed.
Exploring this range, 
from larger to smaller uncertainties,
we discover that $A_M (\veps)$
starts off like the {\em L-A}-integrable circle, 
$A_\infty(\veps) \sim \veps^{-\frac{1}{2}}$,
and successively redirects its course on a different line,
with the same exponent:
a transition from
{\em L-A} to {\em A}-integrability has taken place.

These observations can be derived rigorously from the analysis
developed in the Appendix, with the result:
\beq
   A_M(\veps) \simeq 
    \left\{ 
\begin{minipage}{5.0cm}
$
       \frac{\pi}{2} \frac{1}{ \sqrt{2 \veps}} \;\;  $
      for $  \;\; M \veps \gg H ,
$ \\
$ 
      \frac{\pi}{2} \frac{1}{\sqrt{M \veps}}  $
       for $ \;\; M \veps \ll H ,
$
\end{minipage}
 \right. 
\nuq{forcer}
where the crucial functional dependence on $M$ is
apparent, and where 
$H$ is a constant, which plays a similar r\^ole to
Planck's constant in the ``usual'' quantum mechanics.
In the above equation, the first behaviour is that of the
pure circle (correspondence). 
The second equation in (\ref{forcer}) 
leads to an estimate for the average complexity 
which increases only logarithmically with $N$ and $M$:
\beq
     \Lambda (N)  \leq C + 2 \log_2 N +  \log_2 M,
\nuq{komp1}
where $C$ is a positive constant, 
independent of $M$ and $N$,
and therefore from eq. (\ref{big1}) one obtains \cite{costa} 
\beq
     L (N)  \leq C' + 3 \log_2 N + 2 \log_2 M  ,
\nuq{komp2}
where $C'$ is another positive constant.
This result should be compared with the existing literature on
the {\em topological} complexity of symbolic dynamics
\cite{nothube}.

In conclusion, we see that vertices add logarithmically to 
the complexity, but only as long as $M \veps \ll H$,
{\em i.e.} $\log M < \log H + \Lambda$. 
We have so reconciled the observation of ref. \cite{ford2} and
the established knowledge on polygonal billiards.
We can now turn to a more delicate problem: what happens of
orbital complexity in the polygonal approximations of
a {\em chaotic} billiard.

\section{Orbital Complexity and Correspondence in Chaotic Billiards}

In the previous section we have established that
rational approximations of the circle cannot 
be called chaotic,
not even under the assumption of finite precision. 
Is this achievement possible
to their more sophisticated relatives, those inscribed
in the stadium~? After all, as $M$ tends to infinity,
these fellows tend to a fully chaotic system, and {\em one thing
is certain, correspondence principle must be obeyed}
\cite{corre}. 
\begin{figure}[htbp]
\centering\epsfig{file=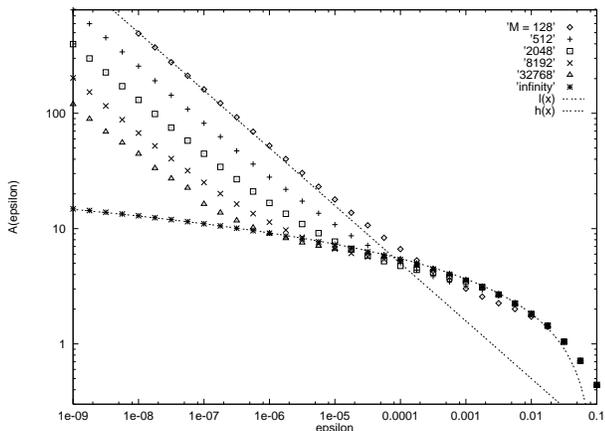,width=0.7\linewidth,angle=270}  
\caption{Average coding length $A_M(\veps)$ vs $\veps$ for 
the quarter stadium and its polygonal quantizations, with $S=2$. 
The values of $M$ are powers of two.
The curves $l(x)$
and $h(x)$ are given in eq. (\ref{forsta})
with $M=\infty$ and $M=128$, respectively, and $A=-1.731$,
$B= 0.831$, $\rho=1.78$. 
}  
\protect\label{stad2.gle}
\end{figure}
Fig. \ref{stad2.gle} is the analogue of Fig. \ref{cer2.gle},
now for the quarter stadium billiard table.
The abscissa is $\veps$ in log scale, i.e. minus
$\Lambda$. The logarithmic character of the curve $A_\infty$
clearly reveals that the complexity of trajectories 
grows linearly with length: chaos is here manifest in its
essence. 
For large values of $M \veps$, $A_M(\veps)$ is 
roughly equal to $A_\infty(\veps)$.
For instance,
over the interval $(10^{-5},10^{-2})$  $A_{32768}$
is a logarithmic function of $\veps$; we can therefore expect that
a corresponding range in $N$ exist so that the average
complexity $K(N)$ grows linearly:
here, trajectories are random, in the
sense explained in Sect. IV. Notice that the existence of this range
is what permits ``sensible'' numerical experiments of chaotic
motion. 

The effect of vertices noticed in the circle 
quantization appears again
when following the curve to the left, as it
leaves $A_\infty$ staying slightly {\em lower}
than this latter,
i.e. showing a relative {\em increase}  in the complexity
of the motion. As noted above, this increase is quantified by the
logarithmic contribution $\log M$.
Alas, this excess of zeal 
rapidly turns into failure: the algorithmic simplicity
of polygonal boundaries is soon detected, and
$A_{32768}$ starts to grow like $\veps^{-\frac{1}{2}}$.
What has happened is that
the complexity estimate (\ref{komp1}), previously dominated by 
that of the curved stadium, takes over for good.
In fact,
an exact analysis can be performed here too, showing that
\beq
   A_M(\veps) \simeq 
    \left\{ 
\begin{minipage}{5.0cm}
$
   A - B \log \veps \;\; $
      for $  \;\; M \veps \gg H ,
$ \\
$ 
 \frac{\pi}{4} \frac{1}{\sqrt{\rho M \veps}} 
\;\;\;\;\;\;\; $
       for $ \;\; M \veps  \ll H ,
$
\end{minipage}
 \right. 
\nuq{forsta}
where $A$, $B$, $\rho$ and $H$ are suitable constants, 
the last playing the same r\^ole as in the previous
section.
Consequently to this equation, in the region $M \veps \ll H$ 
complexity estimates of the form
(\ref{komp1}),
(\ref{komp2}) still apply.

The end of this paper is in sight,
and we must start drawing conclusions.
There is a significant difference between the case of the
circle (Fig. \ref{cer2.gle} and eq. (\ref{forcer}) )
and of the stadium quantization
(Fig. \ref{stad2.gle} and eq. (\ref{forsta})):
while the former, as we noted, shows a transition from {\em A}
to {\em L-A}-integrability, the second 
matches {\em A}-integrability with randomness,
and this mating is troublesome, to say the least. 
Perhaps the most evident consequence
is the following:
let me ask how long
must a symbolic sequence be to perceive the
finiteness of the number of boundary sides
via its algorithmic manifestations.
The answer is instructive.
In the polygonal quantization of the circle,
and of the stadium as well,
the critical value $\veps_M$ at which the $M=\infty$
curve is abandoned scales as $\veps_M \sim \frac{H}{M}$.
Yet, at the same time the different structures of the two problems
imply that the length of polygonal-curved accordance
is proportional to $\sqrt{M}$
in the integrable case, and only to 
the {\em logarithm} of $M$ for the chaotic stadium \cite{zas}.

These results have been obtained by explicit calculation. Yet,
I believe their algorithmic nature 
to be more important:
we must realize that both circle
and stadium polygonal quantizations are dynamical
systems endowed with 
an amount of complexity which scales as the
logarithm of $M$.
Yet, while the $M$-circle dynamics unfolds this complexity
slowly, at a logarithmic pace, its $M$-stadium relative
does it eagerly, linearly in time, so that the
complexity reservoir is quickly exhausted.
In the Conclusions
I shall briefly present my views on the 
physical implications of these facts.

\section{Conclusions: Chaos in Nature, is There Any ?}
Algorithmic complexity theory, used here in a
physicist's fashion via the concept of average first
error time, clarifies
the simplicity of the dynamics of polygonal
billiards, and offers a solution of the paradox
which arises from the juxtaposition of ref. 
\cite{ford1} and \cite{ford2}.
At the same time, this theory provides a complete characterization,
in all ranges of parameters,
of the dynamical properties of the families of billiards studied in this paper.
These latter, in turn, 
are significative representatives of integrable and chaotic 
billiards, and we can safely expect our results to be quite
general.

In essence, the algorithmic estimates derived 
in this paper are manifestations
of the same phenomenon observed in 
the Schr\"odinger quantization of 
the Arnol'd cat 
\cite{berry1,physd,copen,amj} (see eqs. (8) and (9) of ref. \cite{copen}),
of bounded systems \cite{ilg},
and in the classical dynamics of discrete systems
(see Sect IV of ref. \cite{discr}).
Truly then, the nature of chaos and order is
information content~!
Yet, if we admit this, we must be prepared to
bear the consequences when going back
to the problem of quantum mechanics: then, the metaphor
of rational billiards exposed here permits 
to understand the claim that the
{\em correspondence principle} is {\em validated} 
for integrable systems, and 
{\em violated} for chaotic ones. It was shown in the last section,
in fact, that in the first case
a discrete 
nature--in which only polygons are allowed--would 
let us play with our curved theoretical models for a long time.
For the same reason, in the quantum mechanics of
an integrable system, the action of increasing a semi-classical parameter 
provides a complexity reservoir which keeps up with
the classical description for a
time-span which grows appreciably, as a power-law,
in this parameter.
In the second case, to the contrary,  
the time of chaotic freedom is logarithmically short, and
the essence of
correspondence, to regain classical/curved
complexity by quantum/polygonal computations
is exponentially remote in the semi-classical parameter \cite{steven}.
Quantum dynamics is a computable theory, and can ``correspond'' to
the uncomputable classical mechanics only as far as its
algorithmically simple nature allows it \cite{copen2}.

Notwithstanding this evidence,
people have been for a long time
reluctant to admit that there is not such a thing as ``quantum
chaos'', or that this chaos is something different--and less--than
deterministic randomness. This hindrance is now history.
To the contrary, Boris Chirikov has appreciated this fact
since the very beginning, and his notions of transient 
chaos \cite{treotto}
and pseudo-chaos \cite{trenove},
\cite{notchi1},
show this very clearly.
Moreover, he has put a strong emphasis on this latter concept:
largely simplifying, but I believe appropriately,
one could say that Chirikov views the
pseudo-chaos (that we call {\em A}-integrability) of
discrete systems (and by consequence of the digital computers as well)
as a faithful representation of what takes place in
quantum mechanics \cite{guzzibo}--and
in nature, at all levels.
At this very last point we part company:
in fact,
an unsettled debate still holds on the meaning and 
validity of the correspondence principle \cite{jens}, 
and on the question
whether quantum {\em A}-integrability is enough to cope with a world
in which chaos seems to be essential. 
In my works with Joe Ford  a negative answer
to this question  has been presented,
suggesting that--in this sense--quantum mechanics might be 
{\em incomplete}, and that the ``reductionist'' program of 
deriving classical chaos from quantum mechanics could not
be achieved \cite{berry2}. 

Certainly, classical mechanics is inadequate to describe reality,
but it contains the gene of non-computability, which we believe 
should be present in a complete 
theory of nature; at the same time, 
the more fundamental quantum mechanics inherited
this gene in too a tame form to be effective. This is, {\em in nuce},
the point of contention.
Will time settle the matter--or 
will it be that the general indeterminacy
principle foreseen by Ford \cite{jotto} will change everything
around ?

\section{APPENDIX:  Average first error time calculation}

We shall now justify the formulae
presented in Sects. VI and VII via a direct calculation
of the average first error time $A(\veps)$.
Firstly, let us consider the case of the (integrable) circular
billiard. The angle $\phi$ is a constant of the motion:
letting $\omega := \pi - 2 \phi$ the motion is $l_n = l_{n-1}
+ \omega$, the usual ergodic (for irrational $\omega/\pi$)
rotation of the circle.
Trajectories with different initial
angle $\phi$ separate at a linear speed $2 \veps$, where 
$\veps$ has the same meaning as in the main body of the paper
(see Fig. \ref{cerchio}).
Consequently, a bunch of trajectories opens up linearly in 
time like a slightly unfocused light beam.
An {\em error} occurs when a boundary point of the circle partition
which determines the coding
enters this light cone.
Let us now estimate the average time required for this to happen.

Let $S$ to be the number of cells of the partition of the boundary,
and $\delta_S := \frac{2 \pi}{S}$ their common length.
Let also $\theta_n := l_n \; \mbox{mod} \; \delta_S$,
$n = 1,2,\ldots$.
According what we have just said, an error occurs when 
the angle $\theta_n$ of
the reference trajectory (the center of the beam) comes
within $n \tilde{\veps}/2$ of zero or one, where $\tilde{\veps} :=
\frac{4}{\delta_S} \veps$. 
Let $p_n(\veps)$ be the probability that the first 
error occurs at time $n$. Ergodicity implies that 
$p_1(\veps)  = \tilde{\veps}$. 
Certain approximations are required to evaluate $p_n$ for $n >1$:
we assume that $\{\theta_n \}$, $n=1,2,\ldots$  are
uncorrelated random variables, so that
$p_n(\veps)  = n \tilde{\veps} \prod_{j=1}^{n-1} (1 - j \tilde{\veps})$.
Estimating then the asymptotic (small $\tilde{\veps}$)
behaviour of $A(\veps) = \sum_n n p_n(\veps)$ by 
standard techniques gives
$
  {A}({\veps}) \simeq \sqrt{\frac{\pi}{2}} 
  \tilde{\veps}^{- \frac{1}{2}}, 
$
and therefore
\beq
   A(\veps) \simeq \frac{\pi }{2}
  {S}^{- \frac{1}{2}}
  {\veps}^{- \frac{1}{2}},
\nuq{app1}
which well reproduces the
behaviour numerically found in Sect. VI.

Surprisingly, the same $\veps$ dependence can be shown to
hold for the polygonal billiards considered in this paper. 
The presence of the vertices imposes a 
more sophisticated analysis.
When the beam bounces fully on the same
polygonal side, its angular amplitude is conserved,
and its transverse dimension grows linearly in time
on average, exactly as
it does in the circular billiard.
To the contrary, when the beam impinges on a vertex
of amplitude $\alpha$, its angular amplitude
is increased by the amount $2 (\pi - \alpha)$. 
In the polygonal approximations
of the circle, we have $\alpha = (1 - 2/M) \pi$,
while
$\alpha = (1 - 1/2M) \pi$ for the quarter stadium,
so that $\phi_M^{(c)} := \frac{4 \pi}{M}$ 
is the resulting perturbation in the circular case,
and 
$\phi_M^{(s)} := \frac{\pi}{M}$ in the quarter stadium.

Two regimes must now be considered.
Firstly, when $\veps$ is much larger than $\phi_M$,
the effect of vertices is negligible, 
trajectories behave as in the corresponding curved
billiards, and $A_M(\veps)$ coincides to a good approximation
with $A_\infty(\veps)$.
In the opposite regime, when $\phi_M$ is  much larger than
$\veps$, impinging on a vertex causes a major enlargement of
the beam. We can assume that this leads almost immediately
to an error: a similar analysis to that performed in
the circular case can then be carried out.
We let $\delta_M = \frac{2 \pi}{M}$ be the approximate length
of the polygonal sides: this quantity plays here
the same r\^ole $\delta_S$ did above. The result is then
\beq
   A_M(\veps) \simeq \frac{\pi }{2}
  {M}^{- \frac{1}{2}}
  {\veps}^{- \frac{1}{2}},
\nuq{app2}
for the $M$-circle, and
\beq
   A_M(\veps) \simeq \frac{\pi }{4}
   \rho^{- \frac{1}{2}}
  {M}^{- \frac{1}{2}}
  {\veps}^{- \frac{1}{2}}
\nuq{app3}
for the $M$-stadium, where $\rho$ 
is a parameter which measures the effective length
of trajectories in between two coded bounces.
Finally, the intermediate region between the two regimes
is determined by the condition $\phi_M \sim  \veps $ described above,
which becomes the
``indeterminacy principle''
$M \veps \sim H$, where $H$ is a constant.


\begin{thebibliography}{99}
\bibitem{notgam}
This game is not as uncommon as it might seem: 
who has not been challenged, after a good shot,
to {\em do it again if you can~?}
\bibitem{perio} The periodic orbit theory of this function
is interesting, but will 
not be developed in this paper.
\bibitem{costa} 
When computing $K$ from $\Lambda$,
observe that the coding length
$C_{\mbox{\scriptsize boundary}}$ of the set of boundaries can be easily
shown to be logarithmically increasing in $M$.
\bibitem{corre} Capitalized statement of a colleague 
that Joe Ford had provocatively posted on a wall in his office.
\end{thebibliography}
\end{document}